\documentclass[conference]{IEEEtran}

\usepackage{cite
}
\usepackage{graphicx}
\usepackage{algorithm}
\usepackage{algpseudocode}
\usepackage{amsthm}
\usepackage{amssymb}
\usepackage{amsfonts}
\usepackage[cmex10]{amsmath}
\usepackage{color}
\usepackage{comment}

\newcommand{\specialcell}[2][c]{%
  \begin{tabular}[#1]{@{}c@{}}#2\end{tabular}}

\newtheorem{theorem}{Theorem}
\newtheorem{lemma}{Lemma}
\newtheorem{definition}{Definition}

\newcommand{\be}{\begin{equation}}
\newcommand{\ee}{\end{equation}}
\newcommand{\bse}{\begin{subequations}}
\newcommand{\ese}{\end{subequations}}
\def\bee#1\eee{\begin{align}#1\end{align}}
\newcommand{\nnb}{\nonumber}

\ifodd 1
\else

\fi

\ifCLASSINFOpdf
\else
\fi
\begin{document}
%
\title{On the Min-Max-Delay Problem:\\NP-completeness, Algorithm, and Integrality Gap}

\author{\IEEEauthorblockN{Qingyu Liu\IEEEauthorrefmark{1}, Lei Deng\IEEEauthorrefmark{2}, Haibo Zeng\IEEEauthorrefmark{1}, Minghua Chen\IEEEauthorrefmark{2}}
	\IEEEauthorblockA{\IEEEauthorrefmark{1}Department of Electrical and Computer Engineering, Virginia Tech, USA}
	\IEEEauthorblockA{\IEEEauthorrefmark{2}Department of Information Engineering, The Chinese University of Hong Kong, Hong Kong}
	}
\maketitle


\begin{abstract}
We study a delay-sensitive information flow problem where a source streams information to a sink over a directed graph $G\triangleq(V,E)$ at a fixed rate $R$ possibly using multiple paths to minimize the maximum end-to-end delay, denoted as the \textsf{Min-Max-Delay} problem. Transmission over an edge incurs a constant delay within the capacity. We prove that \textsf{Min-Max-Delay} is weakly NP-complete, and demonstrate that it becomes strongly NP-complete if we require integer flow solution.
We propose an optimal pseudo-polynomial time algorithm for \textsf{Min-Max-Delay}, with time complexity $O(\log (Nd_{\max}) (N^5d_{\max}^{2.5})(\log R+N^2d_{\max}\log(N^2d_{\max})))$, where $N \triangleq \max\{|V|,|E|\}$ and $d_{\max}$ is the maximum edge delay.
Besides,  we show that the integrality gap, which is defined as the ratio of the maximum delay of an optimal
integer flow to the maximum delay of an optimal fractional flow, could be arbitrarily large.
\end{abstract}


%

\section{Introduction}
Delay-sensitive network flows have strong applications in many domains, including  communication networks, cyber-physical systems, transportation networks and evacuation planning~\cite{ahuja1993network}. 
In communication networks, video conferencing requires 
the video delivery delay to be no more than $250$ms to ensure a good interactive conferencing experience~\cite{chen2013celerity}, and
packet delivery delay highly affects user experience and hence revenue in data center networks of cloud service providers like Amazon, Microsoft and Google~\cite{alizadeh2010data}.
In cyber-physical systems, shorter delay for control messages can improve control quality~\cite{bai2006towards}. 
In transportation networks, timely delivery is critical to deliver perishable goods~\cite{ashby2006protecting}. 
In evacuation planning, it is important to move all people from hazardous areas to safe areas as soon as possible~\cite{nolan2005storm}.


Theoretically, there are mainly two different delay-sensitive flow models: the flow-amount model and the flow-rate model.
The flow-amount model is suitable to the applications where the flow is generated once,
while the flow-rate model is suitable to the applications where
the flow is generated continuously.
For both models, two different delay-sensitive network flow problems are important: maximizing the flow (amount or rate) subject to a maximum delay constraint, and minimizing the maximum delay subject to a flow (amount or rate) requirement.
We summarize related studies in Tab.~\ref{tab:comparision}.

For the flow-amount model, the first problem, called \emph{dynamic flow problem}~\cite{ford1962flows},
is to maximize the flow amount to be delivered from a source to a sink within a
given time horizon. The authors in~\cite{ford1962flows} show that it can be formulated as a min-cost flow problem and thus solved in polynomial time by various min-cost flow algorithms.
The second problem, called \emph{quickest flow problem}, is to minimize the time horizon to deliver a given amount of flow from
a source to a sink and is also solvable in polynomial time~\cite{lin2015quickest}.

For the flow-rate model, the first problem, called \emph{delay-constrained max-flow problem}~\cite{dong2016traffic}, is to maximize the flow rate
to be sent from a source to a sink while the end-to-end delay is bounded above by a given delay constraint. The problem has been proved to be NP-complete~\cite{dong2016traffic}.
The second problem, called \emph{maximum latency problem}~\cite{correa2004computational}, is to minimize the maximum end-to-end delay that flow units experience from a source to a sink
while satisfying a given flow rate requirement and has been proved to be NP-complete~\cite{correa2004computational}.

\begin{table}[t]
\centering
\caption{Delay-sensitive Network Flow Problems.}
\label{tab:comparision}
\begin{tabular}{|c|c|c|}
\hline
& \specialcell{Max-flow subject to \\ delay constraint} & \specialcell{Min-max delay subject to\\ flow requirement} \\ \hline
Flow-amount Model    & \cite{ford1962flows,ford1956maximal} & \cite{burkard1993quickest,lin2015quickest} \\ \hline
Flow-rate Model  & \cite{wang2014sending,dong2016traffic} & \cite{correa2004computational,correa2007fast}, our paper \\ \hline
\end{tabular}
\end{table}

This paper studies the \textsf{Min-Max-Delay} problem, which is similar to the maximum latency problem~\cite{correa2004computational,correa2007fast} except for two key differences:
\begin{itemize}
\item In our \textsf{Min-Max-Delay} problem, each edge has an integer capacity such that the assigned flow rate cannot exceed the given capacity, while
there is no capacity constraint in the maximum latency problem;
\item Edge delay is an integer in \textsf{Min-Max-Delay}. But it is a flow-dependent function in the maximum latency problem.
\end{itemize}

These two differences can capture many applications that cannot be handled by the maximum latency problem.
For example, in the communication networks, the flow-dependent queueing delay becomes negligible and thus the constant 
propagation delay dominates the edge delay in the light load scenario~\cite{Propagation}.
In the transportation networks, the ground vehicle speed (or equivalently the time to pass the road) remains nearly
constant before reaching a certain flow rate~\cite{chin1991examination}.
Due to these two key differences, existing results on the maximum latency
problem including the complexity analysis and approximation
algorithms~\cite{correa2004computational,correa2007fast}
are not applicable to our problem.

In this paper, we make the following contributions:

$\mbox{\ensuremath{\rhd}}$
We prove that \textsf{Min-Max-Delay} is weakly NP-complete in Thm.~\ref{thm:weak-np-complete-min-max-delay} based on the results in Sec.~\ref{subsec:np-complete-min-max-delay} and Sec.~\ref{sec:algorithm}.

$\mbox{\ensuremath{\rhd}}$
We propose a binary-search algorithm which can find the optimal solution to
\textsf{Min-Max-Delay} in pseudo-polynomial time in Sec.~\ref{sec:algorithm}.
The time complexity of the algorithm is $O(\log (Nd_{\max}) (N^5d_{\max}^{2.5})(\log R+N^2d_{\max}\log(N^2d_{\max})))$, where $N \triangleq \max\{|V|,|E|\}$, $R$ is the rate requirement, and $d_{\max}$ is the maximum edge delay.
The complexity is pseudo-polynomial in the sense that it is polynomial
in the numeric value of the problem input $d_{\max}$,
but is exponential in the bit length of the problem input, i.e., $\log(d_{\max})$~\cite{garey1978strong}.

$\mbox{\ensuremath{\rhd}}$
We prove that \textsf{Min-Max-Delay} becomes strongly NP-complete if each path can only have an integer flow 
in Thm.~\ref{thm:strong-np-complete-integer-min-max-delay} of Sec.~\ref{subsec:strong-np-complete-integer-min-max-delay}.

$\mbox{\ensuremath{\rhd}}$
In Sec.~\ref{sec:counter-example}, we further construct a network to
show that the integrality gap, which is defined as the ratio of the maximum delay of an optimal
integer flow to the maximum delay of an optimal fractional flow, could be arbitrarily large.
This example illustrates additional intriguing difficulties for the
integer version of \textsf{Min-Max-Delay} problem.

\section{System Model and Problem Formulation}\label{sec:definition}
We consider a network modeled as a directed graph $G \triangleq (V,E)$ with $|V|$ nodes
and $|E|$ edges. We define $N\triangleq \max\{|V|,|E|\}$. Each edge $e \in E$ has a non-negative integer capacity $c_e$ and a non-negative integer delay $d_e$.
We define $d_{\max} \triangleq \max_{e \in E} d_e$. A source node $s \in V$ needs to send a positive integer rate $R$ to a sink node $t \in V\backslash\{s\}$.

We denote $P$ as the set of all paths from $s$ to $t$. For any path $p \in P$, we denote its path delay as
$d^p \triangleq \sum_{e\in E:e\in p} d_e$. A flow  solution $f$ is defined as the assigned
flow rate over $P$, i.e., $f \triangleq \{f^p: f^p \ge 0, p \in P\}$. For a flow solution $f$, we define
$f_e \triangleq \sum_{p \in P: e \in p} f^p$ as the flow rate on edge $e \in E$.
We further define the \emph{maximum delay} of a flow solution $f$ as
\be
\label{eqn:maxdelay}
\mathcal{D}(f) \triangleq \max_{p \in P: f^p>0}  d^p,
\ee
i.e., the maximum delay among paths that carry \emph{positive} rates.

We consider the problem of finding a flow solution $f$ to minimize the maximum delay
$\mathcal{D}(f)$ while satisfying both the rate requirement and edge capacity constraints. We denote the problem as \textsf{Min-Max-Delay}. It is formulated as
\bse
\label{eqn:1Problem}
\bee
\min & \quad D \label{eqn:1obj}
\\ \mbox{s.t. } & \quad \sum_{p\in P}f^{p} = R, \label{eqn:1rate}\\
& \quad f_e = \sum_{p \in P:e\in p}f^{p}\leq c_{e},\;\;\forall e\in E, \label{eqn:1capacity}\\
& \quad f^{p}\left(d^p -D\right)\leq0,\;\;\forall p\in P,\label{eqn:1delay}\\
\mbox{vars. } & \quad f^{p}\geq0,\;\;\forall p\in P. \label{equ:1non-negative}
\eee
\ese
where~\eqref{eqn:1obj} together with~\eqref{eqn:1delay} define our objective to minimize the maximum path delay for $s-t$ paths that carry positive rates (called \emph{flow-carrying paths}). Constraint
\eqref{eqn:1rate} restricts that the source $s$ sends $R$ rate to the sink $t$, and constraint
\eqref{eqn:1capacity} requires that the flow rate on edge $e$ does not exceed its capacity $c_e$.

From formulation \eqref{eqn:1Problem} we observe two difficulties to solve
\textsf{Min-Max-Delay}: (i) The number of paths (number of variables) can exponentially increase w.r.t.
the network size, and (ii) formulation~\eqref{eqn:1Problem} is non-convex due to constraint~(\ref{eqn:1delay}).

The integer version of problem \textsf{Min-Max-Delay}, denoted as \textsf{Int-Min-Max-Delay}, further requires that each path carries an integer flow rate, i.e., replacing~\eqref{equ:1non-negative} by
\be
\mbox{vars. } \ \ f^{p} \in \mathbb{Z}^+, \;\;\forall p\in P.
\label{equ:non-negative-integer}
\ee

We let $f_{\textsf{MM}}$ (resp. $f_{\textsf{IMM}}$) to be an optimal flow solution to problem \textsf{Min-Max-Delay}  (resp.
\textsf{Int-Min-Max-Delay}). Then, we define the integrality gap as,
\be \label{equ:def-int-gap}
\textsf{Int-Gap} \triangleq \mathcal{D}(f_{\textsf{IMM}})/\mathcal{D}(f_{\textsf{MM}}),
\ee
which is the maximum delay ratio of the integer flow solution $f_{\textsf{IMM}}$
to the possibly fractional flow solution $f_{\textsf{MM}}$.

\section{NP-completeness}\label{sec:hardness}
In this section, we analyze the computational complexity of our two problems, \textsf{Min-Max-Delay} and \textsf{Int-Min-Max-Delay}.
In Sec.~\ref{subsec:np-complete-min-max-delay}, we prove that \textsf{Min-Max-Delay} is
NP-complete based on the polynomial reduction from the NP-complete partition problem~\cite{garey1978strong}.
In Sec.~\ref{subsec:strong-np-complete-integer-min-max-delay}, we further prove that \textsf{Int-Min-Max-Delay} is
NP-complete in the strong sense
based on the pseudo-polynomial transformation
from the classic strongly NP-complete 3-partition problem~\cite{garey1978strong}.

\subsection{NP-completeness for \textsf{Min-Max-Delay}} \label{subsec:np-complete-min-max-delay}
To analyze the computational complexity of \textsf{Min-Max-Delay}, we first define partition and the partition problem.

\begin{definition}[Partition] \label{def:partition}
	Given a non-empty set $A$, its partition is a set of non-empty subsets such that each element in $A$ is in exactly one of these subsets.
	\label{Def:PartitionSet}
\end{definition}
\begin{definition}[Partition Problem~\cite{garey1978strong}]  \label{def:partition-problem}
Given a set of $n$  positive integers $A= \{ a_1,a_2,...,a_n \}$ with sum $ \sum_{a_i \in A} a_i=2b$.
Is there a partition $\{A_1,A_2\}$ of $A$
such that $ \sum_{a_i\in A_1} a_i=\sum_{a_j\in A_2} a_j= b$?
\end{definition}
\begin{figure}[t!]
	\centering
	\includegraphics[width=1\columnwidth]{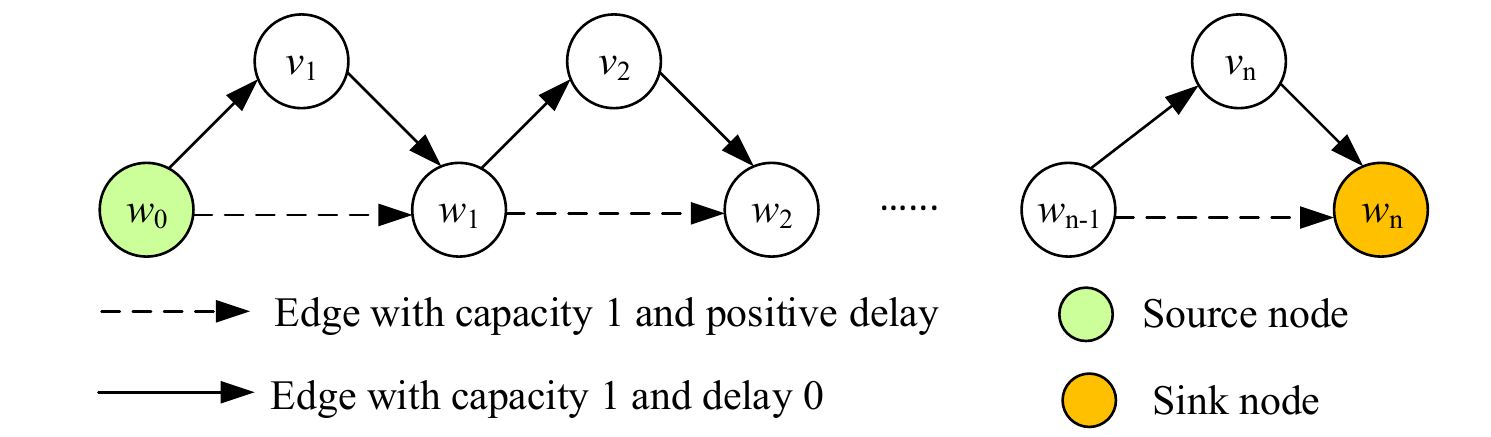}
	\caption{Reduced network graph from partition problem.}
	\label{fig:2partition}
\end{figure}

The partition problem is known to be NP-complete~\cite{garey1978strong} (in the weak sense). We now leverage it to
prove that \textsf{Min-Max-Delay} is NP-complete.

\begin{theorem}\label{thm:np-complete-min-max-delay}
The decision version of \textsf{Min-Max-Delay} problem is NP-complete.
\end{theorem}
\begin{IEEEproof}
For any partition problem we
construct a graph $G'$ with $(2n+1)$ nodes and $3n$ edges as in Fig.~\ref{fig:2partition}.
All edges have unit capacity.
Each edge $(w_{i-1}, w_{i})$ has a delay of $a_i$ for all $i=1, \cdots, n$,
while edges $(w_{i-1}, v_i)$ and $(v_i, w_i)$ have a delay of zero. Obviously it takes polynomial time to construct the graph $G'$.
We then consider the following decision problem of \textsf{Min-Max-Delay}:
for graph $G'$ with source $s=w_0$, sink $t=w_n$, and
flow rate requirement $R=2$,
is there any feasible flow $f$ such that the maximum delay $\mathcal{D}(f) \le b$?

Now we prove the partition problem answers ``Yes" if and only if the decision version of \textsf{Min-Max-Delay} answers ``Yes".

\textbf{If Part.}
If the decision problem of \textsf{Min-Max-Delay} answers ``Yes", then there exists a flow $f$ such that $\mathcal{D}(f) \le b$. Since $f$ is feasible, the total rate from $w_0$ to $w_n$ in $f$
is $R=2$. Now due to the capacity constraint and flow conservation,
all edges must exactly have a flow rate 1 to satisfy the requirement $R=2$. The total delay in flow $f$
is
\be
\sum_{p \in P} f^p d^p = \sum_{e \in E} f_e d_e = \sum_{e \in E} 1 \cdot d_e = \sum_{i=1}^{n} a_i = 2b.
\ee

Since $\mathcal{D}(f) \le b$, we have
\be
d^p \le \mathcal{D}(f) \le b, \forall p\in P \text{ with } f^p > 0.
\ee
Also, because the total flow rate is equal to $2$, we have
\be\label{eqn:minmax-totaldelay}
2b = \sum_{p \in P} f^p d^p = \sum_{p \in P: f^p > 0} f^p d^p \le b \cdot \sum_{p \in P: f^p > 0} f^p= 2b.
\ee
As both ends in \eqref{eqn:minmax-totaldelay} are the same, it must be
\be
d^p = b, \forall p \in P \text{ with } f^p > 0.
\ee
Therefore, all flow-carrying paths have a path delay of $b$. We choose an arbitrary flow-carrying path $p$.
Since all solid edges have a delay of 0, the path delay of $p$ is the delay of all dashed edges that belongs to
$p$. We consider the set $A_1$ that contains $a_i$ if edge $(w_{i-1}, w_i) \in p$. Clearly, it holds that $\sum_{a_i \in A_1} a_i =b$. We then define $A_2 = A \backslash A_1$. It shall be
$\sum_{a_j \in A_2} = \sum_{a_k \in A} - \sum_{a_i \in A_1} = 2b - b = b$. $A_1$ and $A_2$ are thus
a partition of set $A$ and meet the requirement of the partition problem. Hence, the partition problem answers ``Yes".

\textbf{Only If Part.}
If the partition problem answers ``Yes", then there exists a partition $A_1$ and $A_2$ such that $\sum_{a_i \in A_1} a_i = \sum_{a_j \in A_2} a_j = b$.
We now construct two paths $p_1$ and $p_2$.
\begin{itemize}
\item $\forall i \in [1,n]$, if $a_i \in A_1$, we put edge $(w_{i-1}, w_i)$ into path $p_1$; otherwise, we put
$(w_{i-1}, v_i)$ and $(v_i, w_i)$ into $p_1$.
\item $\forall i \in [1,n]$, if $a_i \in A_2$, we put $(w_{i-1}, w_i)$ into $p_2$; otherwise, we put $(w_{i-1}, v_i)$ and $(v_i, w_i)$ into $p_2$.
\end{itemize}
Due to the definition of a partition (see Definition~\ref{def:partition}), $A_1$ and $A_2$ are two disjoint sets, i.e.,
$A_1 \cap A_2 = \emptyset$. Thus, we can easily see that $p_1$ and $p_2$ are two disjoint $s-t$ paths, i.e.,
$p_1$ and $p_2$ do not share any common edge.
We can see that $d^{p_1} = \sum_{a_i \in A_1} = b$, and $d^{p_2} = \sum_{a_i \in A_2} = b$.
We then construct the flow $f$ with only two flow-carrying paths $p_1$ and $p_2$ and set $f^{p_1} = f^{p_2} = 1$.
Since $p_1$ and $p_2$ are disjoint, the capacity constraint is satisfied. Also, since $f^{p_1}+f^{p_2}=R=2$,
the rate requirement is satisfied. Thus $f$ is a feasible flow with maximum delay $\mathcal{D}(f) = b$. Therefore,
the decision problem of \textsf{Min-Max-Delay} answers ``Yes".

Since the partition problem is NP-complete~\cite{garey1978strong} and the reduction can be done in polynomial time,
the decision problem of \textsf{Min-Max-Delay} is also NP-complete.
\end{IEEEproof}

Due to the NP-completeness, it is impossible to solve \textsf{Min-Max-Delay} optimally in polynomial time unless $\text{P} = \text{NP}$.
Later in Sec.~\ref{sec:algorithm}, a pseudo-polynomial time algorithm is proposed to solve \textsf{Min-Max-Delay} optimally,
which further proves that \textsf{Min-Max-Delay} is NP-complete in the weak sense.

\subsection{Strong NP-completeness for \textsf{Int-Min-Max-Delay}} \label{subsec:strong-np-complete-integer-min-max-delay}
We now analyze the computational complexity of \textsf{Int-Min-Max-Delay} where only integer flows are allowed.
We begin with the definition of the 3-Partition Problem~\cite{garey1978strong}.

\begin{definition}[3-Partition Problem~\cite{garey1978strong}] \label{def:3-partition-problem}
Given a set of $n=3k\ (k>0)$ positive integers $A= \{ a_1,a_2,...,a_n \}$
with sum $\sum_{a_i \in A}a_i=k b$ and $b/4<a_i<b/2$ for each element
$a_i$, is there a partition $\{A_1,A_2,...,A_k\}$ of set $A$ such that for each subset $A_j$, $ \sum_{a_i\in A_j}a_i=b$?
\end{definition}

The 3-partition problem has been proved to be NP-complete in the strong sense~\cite{garey1978strong}, which can
be utilized to prove the strong NP-completeness of our problem \textsf{Int-Min-Max-Delay}.

\begin{figure}[t!]
	\centering
	\includegraphics[width=1\columnwidth]{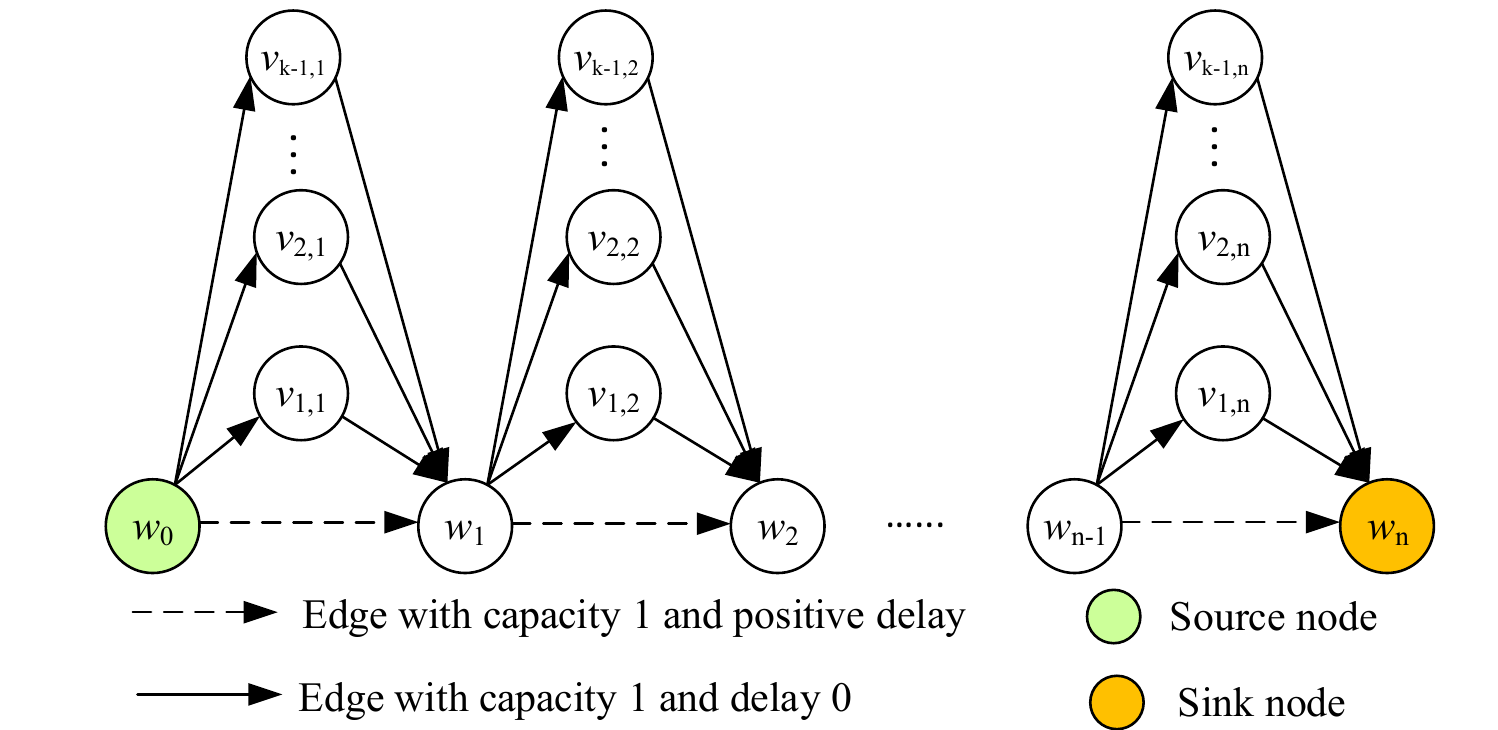}
	\caption{Reduced network graph from 3-partition problem.}
	\label{fig:3partition}
\end{figure}

\begin{theorem}\label{thm:strong-np-complete-integer-min-max-delay}
The decision version of \textsf{Int-Min-Max-Delay} problem is NP-complete in the strong sense.
\end{theorem}

\begin{IEEEproof}
For any 3-partition problem, we construct a network following Fig.~\ref{fig:3partition}. Edge capacity and delay are similarly defined as in the proof to Thm.~\ref{thm:np-complete-min-max-delay}. We set
the flow requirement $R=k$ and thus get the problem \textsf{Int-Min-Max-Delay}.
Such a reduction is a pseudo-polynomial transformation~\cite{garey1978strong}.
Similar to the proof in Thm.~\ref{thm:np-complete-min-max-delay}, we show that
the 3-partition problem answers ``Yes" if and only if the decision problem of \textsf{Int-Min-Max-Delay} answers ``Yes".
Due to the space limit, we put the detailed proof in our technical report \cite{TechnicalReport}.
\end{IEEEproof}

As we will shown later in Thm.~\ref{thm:weak-np-complete-min-max-delay}, \textsf{Min-Max-Delay} is weakly NP-complete.
Thus, Thm.~\ref{thm:strong-np-complete-integer-min-max-delay} shows that
\textsf{Int-Min-Max-Delay} is more difficult than  \textsf{Min-Max-Delay}.
Later in Sec.~\ref{sec:counter-example} we show that the integrality gap defined in \eqref{equ:def-int-gap}
could be arbitrarily large, which further illustrates additional intriguing difficulties for the
\textsf{Int-Min-Max-Delay} problem.

\section{Optimal Pseudo-polynomial Time Algorithm}\label{sec:algorithm}
In this section we propose a pseudo-polynomial time algorithm
to solve \textsf{Min-Max-Delay} optimally.
This, combined with the NP-completeness result in Thm.~\ref{thm:np-complete-min-max-delay}, shows that \textsf{Min-Max-Delay} is NP-complete in the weak sense.


A closely-related problem to \textsf{Min-Max-Delay} is the Delay-Constrained Maximum Flow
problem~\cite{wang2014sending}, denoted as \textsf{DC-Max-Flow}: for the same graph $G$
and a given deadline constraint $T$, finding the maximum flow (rate) such that the delay of any flow-carrying
path does not exceed $T$.
Let us denote $P^T$ as the set of all $s-t$ paths whose path delay does not exceed $T$.
Then \textsf{DC-Max-Flow} can be formulated as
\bse
\label{equ:delay-max-flow}
\bee
\max & \quad \sum_{p \in P^T} f^p, \\
\mbox{s.t.} & \quad f_e = \sum_{p \in P^T: e \in p} f^p \le c_e, \;\; \forall e \in E \\
\mbox{vars.} & \quad f^p \ge 0, \;\; \forall p \in P^T.
\eee
\ese

It is shown that \textsf{DC-Max-Flow} can be solved with an
edge-based flow formulation in pseudo-polynomial time~\cite{wang2014sending}. They implicitly
use the idea of time-expanded graph~\cite{mahjoub2010max} by converting a delay-constrained max-flow problem
in the original graph into a delay-unconstrained max-flow problem in the expanded graph.
In~\cite{wang2014sending}, the authors only consider unit-delay edges, but it is easy to generalize
their result to integer-delay edges. Due to the space limit, we omit the procedure of constructing the expanded graph and directly give the equivalent edge-based flow formulation for \textsf{DC-Max-Flow} as follows~\cite[Proposition 1]{wang2014sending},
\bse \label{eqn:delay-max-flow-hop}
\bee
\max  & \quad \sum_{e \in \textsf{In}(t)} \sum_{d=0}^{T} f_e^{(d)}  \label{eqn:delay-max-flow-obj} \\
\mbox{s.t.}
& \quad \sum_{e \in \textsf{Out}(s)} f_e^{(d_e)}=\sum_{e \in \textsf{In}(t)} \sum_{d=0}^{T} f_e^{(d)}, \label{eqn: total-flow-rate}\\ 
& \quad \sum_{e \in \textsf{In}(v)} f_e^{(d)} = \sum_{e\in \textsf{Out}(v)} f_e^{(d+d_e)},  \nnb \\
& \qquad \qquad \forall v \in V\backslash\{s,t\}, d \in [0,T] \label{eqn:delay-max-flow-conservation} \\
& \quad \sum_{d=0}^{T} f_e^{(d)} \leq c_e, \quad \forall e\in E \label{eqn:delay-max-flow-capacity} \\
\mbox{vars.} & \quad f_{e}^{(d)} \geq 0,  \quad \forall e\in E, d \in [0,T] \label{eqn:delay-max-flow-nonnegative}
\eee
\ese
where $\textsf{In}(v) \triangleq \{e=(w,v): e \in E,w\in V\}$
is the set of incoming edges of node $v$,
$\textsf{Out}(v) \triangleq \{e=(v,w): e \in E, w\in V\}$
is the set of outgoing edges of node $v$,
and $f_e^{(d)}$ is the total flow rate that experiences a delay of $d$ after passing edge $e$ from the source $s$.
The objective \eqref{eqn:delay-max-flow-obj} is the total flow rate that
arrives at the sink $t$ within the delay bound $T$. Eqs.~\eqref{eqn: total-flow-rate} requires the rate entering the network should equal to the rate leaving the network.
Eqs.~\eqref{eqn:delay-max-flow-conservation} are the flow conservation
constraints in the expanded graph. Note that by convention, for any edge $e \in E$, we
set $f_e^{(d)} = 0$ for $d < 0$ and $d > T$.
Eqs.~\eqref{eqn:delay-max-flow-capacity} are the edge capacity constraints.

We now show the relationship between our problem \textsf{Min-Max-Delay} and problem \textsf{DC-Max-Flow}.
For a graph $G$, denote $d^*(R)$ as the minimum maximum delay with rate requirement $R$
(the optimal value of \textsf{Min-Max-Delay}).
For the same graph $G$ and a non-negative integer $T$,
denote $r^*(T)$ as the maximum flow subject to a delay constraint $T$
(the optimal value of \textsf{DC-Max-Flow}).
We have the following lemma.

\begin{lemma} \label{lem:two-problem-relation}
$d^*(R) \le T$ if and only if $r^*(T) \ge R$.
\end{lemma}
\begin{IEEEproof}
\textbf{If Part.} If $r^*(T) \ge R$, then there exists a flow solution over $P^T$ such that $\sum_{p \in P^T} f^p \ge R$. We can thus decrease
the flow solution $f$ to construct another flow solution $\tilde{f}$ such that $\sum_{p \in P^T} \tilde{f}^p = R$.
Since $f$ satisfies the capacity constraints, $\tilde{f}$ must also satisfy the capacity constraints.
Thus, $\tilde{f}$ is a feasible solution to \textsf{Min-Max-Delay} with rate requirement $R$.
In addition, since all flow-carrying paths in $\tilde{f}$ belong to the set $P^T$, we have
$d^*(R) \le \mathcal{D}(\tilde{f}) \le T$.

\textbf{Only If Part.} If $d^*(R) \le T$, then there exists a flow solution $f$ where
the path delay of any flow-carrying path does not exceed $T$. Thus all flow-carrying paths belong to
$P^T$ and $f$ is also a feasible solution to \textsf{DC-Max-Flow} with a delay bound $T$. Thus,
$r^*(T) \ge \sum_{p \in P^T} f^p = R$.
\end{IEEEproof}

Lem.~\ref{lem:two-problem-relation} suggests a binary-search algorithm to solve \textsf{Min-Max-Delay} optimally. Given a lower bound $T_l$ ($=0$ initially) and an upper bound $T_u$ ($=|E|d_{\max}$ initially) of the optimal maximum delay,
in each iteration we solve problem \eqref{eqn:delay-max-flow-hop} with $T=\lceil(T_l+T_u)/2\rceil$.
We then compare $r^*(T)$ with the rate requirement $R$.
If $r^*(T) \ge R$, we update the upper bound as $T_u=T$. Otherwise we update the lower bound
as $T_l = T+1$. The algorithm terminates when $T_l \ge T_u$.

\begin{theorem} \label{the:binary-search-complexity}
The binary search scheme solves \textsf{Min-Max-Delay} problem optimally
and has a pseudo-polynomial time complexity $O(\log (Nd_{\max}) (N^5d_{\max}^{2.5})(\log R+N^2d_{\max}\log(N^2d_{\max})))$.
\end{theorem}
\begin{IEEEproof}
The optimality of the binary search scheme directly follows from Lem.~\ref{lem:two-problem-relation}.
For the time complexity analysis, please see our technical report in \cite{TechnicalReport}.
\end{IEEEproof}

The time complexity shown above is pseudo-polynomial in the sense that it is polynomial in the numeric value of the input $d_{\max}$, but 
it is exponential in the bit length of the $d_{\max}$. Thus, our binary search algorithm could still have high complexity for the 
graph with large $d_{\max}$.
To further reduce the complexity, we adapt the \emph{rounding and scaling} approach~\cite{sahni1977general} and design a fully polynomial time approximation scheme (\textsf{FPTAS}) for our problem \textsf{Min-Max-Delay}.
For any $\epsilon > 0$, our proposed \textsf{FPTAS} can find a $(1+\epsilon)$-approximate solution and the time complexity is polynomial in both problem size and $1/\epsilon$. Due to the space limit, the detailed \textsf{FPTAS} design is shown in our technical report \cite{TechnicalReport}.

A direct result of Thm.~\ref{the:binary-search-complexity} is as follows.
\begin{theorem} \label{thm:weak-np-complete-min-max-delay}
The decision version of \textsf{Min-Max-Delay} is NP-complete in the weak sense.
\end{theorem}
\begin{IEEEproof}
It follows from Thm.~\ref{thm:np-complete-min-max-delay}
and Thm.~\ref{the:binary-search-complexity}.
\end{IEEEproof}


\section{Integrality Gap} \label{sec:counter-example}
We have proved that \textsf{Min-Max-Delay} allowing fractional flow is NP-complete in the weak sense,
but \textsf{Int-Min-Max-Delay} only allowing integer flow solution
is NP-complete in the strong sense. Thus, there are some fundamental differences between these two problems.
To elaborate their differences, in this section, we construct a network to show that the integrality gap defined in \eqref{equ:def-int-gap} could
be arbitrarily large.

\begin{figure}[t]
  \centering
  \includegraphics[width=0.9\linewidth]{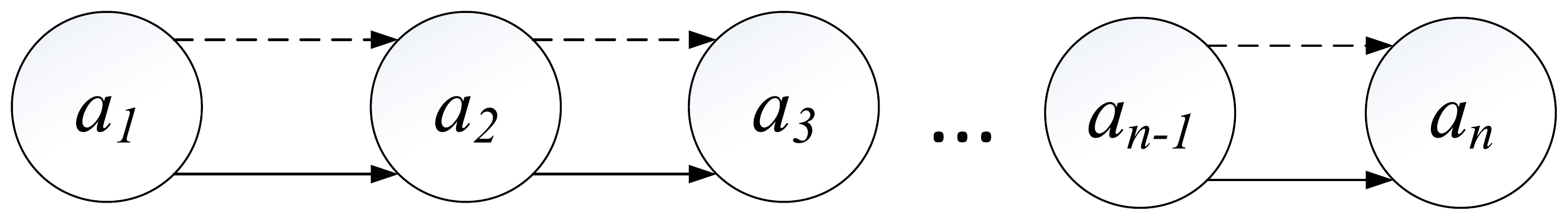}\\
  \vspace{-1ex}
  \caption{A building-block graph with $n \ge 3$ 
  to show that the integrality gap could be arbitrarily large. All edges have unit capacity. Each
  	upper dashed edge has a unit delay while each lower solid edge has a delay
  	of zero.}\label{fig:counter-example}
\end{figure}

We will use the network in Fig.~\ref{fig:counter-example} as a building block to construct the network with  arbitrarily large integrality gap.
Towards that end, for the building-block graph in Fig.~\ref{fig:counter-example} with source node $a_1$ and sink node $a_n$,
we first analyze the maximum delay of problem \textsf{Int-Min-Max-Delay} and problem  \textsf{Min-Max-Delay} 
in Lem.~\ref{lem:IntFlowMM} and Lem.~\ref{lem:FlowMM}, respectively.

\begin{lemma}
	\label{lem:IntFlowMM}
	The \textsf{Int-Min-Max-Delay} flow in Fig.~\ref{fig:counter-example} given $R=2$ has a maximum delay of $\lceil\frac{n-1}{2}\rceil$.
\end{lemma}

\begin{IEEEproof}
Because of the unit capacity constraint for each edge, any flow-carrying path will be assigned a unit flow rate in the integer flow from $a_1$ to $a_n$.
	
	Because of the rate requirement $R=2$, there are exactly two flow-carrying paths in the \textsf{Int-Min-Max-Delay} flow and each of them carries a unit flow rate.
	
	Moreover, because of the capacity constraint, the two flow-carrying paths must be disjoint, namely they share no edges.
	
	Therefore, since the \textsf{Int-Min-Max-Delay} flow minimizes the maximum flow-carrying path delay among all integer flows, the two flow-carrying paths will have path delay of $\lfloor(n-1)/2\rfloor$ and $\lceil(n-1)/2\rceil$ respectively in the \textsf{Int-Min-Max-Delay} flow, leading to a maximum delay of $\lceil(n-1)/2\rceil$.
\end{IEEEproof}

\begin{lemma}
	\label{lem:FlowMM}
	The \textsf{Min-Max-Delay} flow in Fig.~\ref{fig:counter-example} given $R=\frac{n-1}{n-2}$ has a maximum delay of $1$.
\end{lemma}

\begin{IEEEproof}
	First it is straightforward that the \textsf{Min-Max-Delay} flow has a maximum delay no smaller than $1$ since $R>1$.
	
	Next, we will explicitly construct a feasible flow with unit maximum delay: Note that there are $(n-1)$ different $a_1-a_n$ paths containing exactly one dashed edge. We then place $1/(n-2)$ flow rate on each of these $(n-1)$ paths with the rate requirement $R=(n-1)/(n-2)$ satisfied. It is easy to verify that edge capacity constraints are satisfied, too.
\end{IEEEproof}

Note that even though our system model in Sec.~\ref{sec:definition} requires an integer rate $R$, 
problem \textsf{Min-Max-Delay} is also well defined for fractional rate requirement $R$, which is the case in Lem.~\ref{lem:FlowMM}.
However,  later in our constructed graph to show that the integrality gap can be arbitrarily large (see the proof in Thm.~\ref{thm:infinitelargegap}), we set integer flow rate requirement $R$, which is in line with our system model in Sec.~\ref{sec:definition}.

Based on Lem.~\ref{lem:IntFlowMM} and Lem.~\ref{lem:FlowMM}, we use Fig.~\ref{fig:counter-example} as a building-block 
to construct a network with integer edge delay, integer capacity, integer rate $R$ and arbitrarily large \textsf{Int-Gap}.

\begin{theorem}
	\label{thm:infinitelargegap}
	There exists a \textsf{Min-Max-Delay} problem instance such that \textsf{Int-Gap} could be arbitrarily large.
\end{theorem}

\begin{IEEEproof}
	We place $(n-2)$ subnetworks with the topology of Fig.~\ref{fig:counter-example} side by side disjointly and then connect each of them to both a single source $s$ and a single sink $t$. All the outgoing edges of $s$ and the incoming edges of $t$ have a capacity of $2$ and a delay of $0$. We set the rate requirement $R=n-1$.
	
	First, considering Lem.~\ref{lem:FlowMM}, the \textsf{Min-Max-Delay} flow has a maximum delay of $1$ because we can route $(n-1)/(n-2)$ flow rate to each of the $(n-2)$ subnetwork.
	
	Next since the rate requirement $R=n-1$ is strictly larger than $n-2$ which is the number of subnetworks, for any feasible integer flow, there will be at least one subnetwork who is assigned a flow rate of $2$. According to Lem.~\ref{lem:IntFlowMM}, corresponding minimal maximum delay to pass the subnetwork is $\lceil(n-1)/2\rceil$. Thus, the \textsf{Int-Min-Max-Delay} flow will have a maximum delay of $\lceil(n-1)/2\rceil$.
	
	Overall, the \textsf{Int-Gap} is $\lceil(n-1)/2\rceil$. Since $n$ can be arbitrarily large, the gap can be arbitrarily large.
\end{IEEEproof}

Comparing $f_{\textsf{IMM}}$ to $f_{\textsf{MM}}$ under the same rate $R$, \textsf{Int-Gap} can be infinitely large according to Thm.~\ref{thm:infinitelargegap}. However, for any $\epsilon\in(0,1)$, \textsf{Int-Gap} is in fact upper bounded by $1/\epsilon$ if we instead compare $f_{\textsf{IMM}}$ with a smaller rate $(1-\epsilon)R$ to $f_{\textsf{MM}}$ with the full rate $R$, as proved in our report~\cite{TechnicalReport}. Overall, \textsf{Int-Min-Max-Delay} is different from \textsf{Min-Max-Delay}.
Though we have designed an optimal pseudo-polynomial time  algorithm and an \textsf{FPTAS} for \textsf{Min-Max-Delay},
we need to tackle some additional intriguing difficulties to design efficient algorithms for the strongly NP-complete problem \textsf{Int-Min-Max-Delay}.


\section{Conclusion}\label{sec:conclusion}
We study the \textsf{Min-Max-Delay} which minimizes the maximum end-to-end delay under a rate requirement in a single-unicast scenario. We prove it is NP-complete in the weak sense, and propose a pseudo-polynomial time algorithm to find its optima. We also show that if integer flow solution is required, the problem becomes NP-complete in the strong sense. 
We further construct a network to show that the integrality gap, which is the maximum delay ratio of the optimal 
integer flow to the optimal fractional flow, could be arbitrarily large.





%
\bibliographystyle{IEEEtran}
\bibliography{References}

\end{document}